\documentclass[aps, pra, floatfix, reprint]{revtex4-2}
\usepackage[dvips]{graphicx}
\usepackage[english]{babel}
\usepackage{amsmath}
\usepackage{amssymb}
\usepackage{bm}
\usepackage{color}
\usepackage{subfigure}
\usepackage{soul, xcolor}
\setstcolor{red}
\usepackage[colorlinks, citecolor = black, linkcolor = black, urlcolor = blue]{hyperref}

\newcommand{\bk}{\mathbf{k}}
\newcommand{\bb}{\mathbf{k}}
\newcommand{\bq}{\mathbf{q}}
\newcommand{\nn}{\nonumber}
\newcommand{\p}{\partial}
\newcommand{\ak}{a_\mathbf{k}}
\newcommand{\amk}{a_{-\mathbf{k}}}

\newcommand{\hh}{\hat{H}}
\newcommand{\tr}{\mathrm{Tr}}
\newcommand{\uk}{u_{\mathbf{k}}}
\newcommand{\vk}{v_{\mathbf{k}}}
\newcommand{\ha}{\hat{\alpha}_{\mathbf{k}}}
\newcommand{\hma}{\hat{\alpha}_{-\mathbf{k}}}
\newcommand{\be}{\begin{eqnarray}}
\newcommand{\ee}{\end{eqnarray}}
\newcommand{\bn}{\mathbf{n}}

\def\ket#1{|#1\rangle}
\def\bra#1{\langle #1 |}
\def\ep#1{\langle #1 \rangle}
\def\rm#1{\mathrm{#1}}
\def\bf#1{\mathbf{#1}}

\begin{document}

\title{Uhlmann Phase Winding in Bose-Einstein Condensates at Finite Temperature}

\author{Chang-Yan Wang}
\email{changyanwang@tsinghua.edu.cn}
\affiliation{Institute for Advanced Study, Tsinghua University, Beijing 100084, China}
\author{Yan He}
\email{heyan\_ctp@scu.edu.cn}
\affiliation{College of Physics, Sichuan University, Chengdu, Sichuan 610064, China}

\begin{abstract}
  We investigate the Uhlmann phase, a generalization of the celebrated Berry phase, for Bose-Einstein condensates (BECs) at finite temperature. The Uhlmann phase characterizes topological properties of mixed states, in contrast to the Berry phase which is defined for pure states at zero temperature. Using the $SU(1,1)$ symmetry of the Bogoliubov Hamiltonian, we derive a general formula for the Uhlmann phase of BECs. Numerical calculations reveal that the Uhlmann phase can differ from the Berry phase in the zero-temperature limit, contrary to previous studies. As the temperature increases, the Uhlmann phase exhibits a winding behavior, and we relate the total winding degree to the Berry phase. This winding indicates that the Uhlmann phase takes values on a Riemann surface. Furthermore, we propose an experimental scheme to measure the Uhlmann phase of BECs by purifying the density matrix using an atomic interferometer. 
\end{abstract}
\maketitle

\section{Introduction}

The Berry phase \cite{berry_quantal_1984, zak_berry_1989}, acquired by a quantum system undergoing an adiabatic evolution, can characterize the topological properties of quantum systems at zero temperature, such as the quantum Hall effect \cite{thouless_quantized_1982}. However, Berry phase is only defined for pure states, while in reality, most quantum systems are in contact with an environment, existing at finite temperatures and consequently described by mixed states. To bridge this gap, the Uhlmann phase was introduced, extending the concept of geometric phase to mixed states through the adiabatic evolution of density matrix \cite{mera_uhlmann_2017, uhlmann_parallel_1986, uhlmann_gauge_1991, uhlmann_density_1993}.

The Uhlmann phase has been the subject of extensive studies for mixed states \cite{ericsson_mixed_2003, ericsson_generalization_2003, kiselev_interferometric_2018, singh_geometric_2003, tidstrom_uhlmann_2003, tong_kinematic_2004}, and has recently attracted considerable attention for investigating the topological properties of various quantum systems at finite temperatures \cite{wang_uhlmann_2023, carollo_uhlmann_2018, gao_quantum_2023, gao_quantum_2023, guo_dynamic_2020, he_thermal_2018, he_uhlmann_2022, hou_ubiquity_2020, mera_dynamical_2018, pi_proxy_2022, villavicencio_uhlmann_2021, viyuela_symmetryprotected_2015, zhang_comparison_2021}, including fermionic lattice models \cite{huang_topological_2014, viyuela_uhlmann_2014, viyuela_twodimensional_2014} and spin systems \cite{hou_finitetemperature_2021, morachisgalindo_topological_2021, villavicencio_thermal_2023}. Despite this, the Uhlmann phase for Bose-Einstein condensates (BECs)—one of the most important states of matter in ultracold atomic systems—has not been explored. Given the high controllability of BEC systems \cite{chin_feshbach_2010}, the study of the Uhlmann phase of BEC can enrich our understanding of topological properties at finite temperatures and for mixed states.

In this paper, we study the Uhlmann phase of BEC by exploiting the so-called $SU(1,1)$ symmetry of the Bogoliubov Hamiltonian. The $SU(1,1)$ group, a non-compact Lie group, includes the transformations preserving the canonical commutation relation of bosonic operators of BEC \cite{perelomov_generalized_1986}. This group has been widely used in quantum optics \cite{puri_mathematical_2001, caves_new_1985}, and recently finds its application in studying the dynamics and quantum echo of BEC \cite{chen_manybody_2020, cheng_manybody_2021, lv_su_2020, lyu_geometrizing_2020, zhang_quantum_2022, zhang_periodically_2020}. Meanwhile, the similar symmetric approach has been generalized to two-component BEC \cite{wang_quantum_2022, wang_quantum_2024, penna_twospecies_2017, richaud_quantum_2017, charalambous_control_2020}. 

By using the $SU(1,1)$ symmetry, the density matrix of the BEC at finite temperature can be parameterized as a Poincare disk, akin to the Bloch sphere for the spin system. 
The adiabatic Uhlmann process is then represented by a trajectory within this disk. By considering different paths, we find that the Uhlmann phase of BEC can differ from the zero-temperature Berry phase, contrary to previous studies on the Uhlmann phase. Meanwhile, we find that the Uhlmann phase of BEC displays a unique winding behavior when the temperature increases from low to high. More interestingly, the total winding degree is equal to the negative of the Berry phase at zero temperature. This winding behavior indicates that the Uhlmann phase takes values in a Riemann surface. We further propose a scheme to measure the Uhlmann phase of BEC experimentally based on purification of the density matrix using an atomic interferometer which uses the interference of atoms to perform precise measurements \cite{cronin_optics_2009}. 

Our work highlights the systematic use of the nontrivial $SU(1,1)$ symmetry of BEC and unveil a novel Uhlmann phase winding feature previously unreported. This winding behavior establishes an interesting link between the geometric phase behaviors at zero and finite temperatures. Moreover, our study also relates this winding behavior to the geometric nature of the Uhlmann phase by demonstrating its association with a curve in Riemann surface.

The rest of this paper is organized as follows: in Sec. \ref{sec:ham}, we review Bogoliubov Hamiltonian of BEC and its $SU(1,1)$ symmetry; in Sec. \ref{sec:uhl}, we briefly review the definition of Uhlmann phase; in Sec. \ref{sec:uhl_bec}, we derive the Uhlmann phase of BEC, and present the general method to numerically calculate the Uhlmann phase of BEC; in Sec. \ref{sec:uhl_wind}, we present the numerical results, and reveal the Uhlmann phase winding; in Sec. \ref{sec:experiment}, we discuss how \st{to} one can observe the Uhlmann phase of BEC experimentally; we conclude in Sec. \ref{sec:conclusion}.

\section{BEC Hamiltonian and $SU(1,1)$ symmetry} \label{sec:ham}

We consider an interacting bosonic gas, the Hamiltonian of which in momentum space can be written as
\begin{eqnarray}
  \hat{H} = \sum_\bk \epsilon_\bk a_\bk^\dag a_\bk + \frac{g}{2V} \sum_{\bk,\bk',\bq} a_{\bk + \bq}^\dag a_{\bk' - \bq}^\dag a_{\bk'} a_{\bk},
\end{eqnarray}
where $a_\bb (a_\bb^\dag)$ is the annihilation (creation) operator for bosons with momentum $\bb$, and $\epsilon_\bb = \bb^2/2m - \mu$, $g = 4\pi \hbar^2 a_s/m$, with $a_s$ the s-wave scattering length, which can be tuned through Feshbach resonance \cite{chin_feshbach_2010}. In the following we will set $\hbar = 1$ for simplicity. 

Due to Bose-Einstein condensation, most bosons will stay at the lowest energy state with $\bb=0$. Making use of the Bogoliubov approximation, we can expand the interaction terms around the condensate up to the quadratic terms of boson operators. Then we arrive at the Bogoliubov Hamiltonian of BEC as follows $\hh_{\mathrm{Bog.}} = \sum_{\bk\neq \mathbf{0}} \hh_\bb + \mathrm{const.}$, where
\begin{eqnarray}
  \hh_\bb = \sum_{i=0}^2 \xi_i(\bk) K_i.\label{ham}
\end{eqnarray}
Here, the coefficients are $\xi_0 = 2(\epsilon_\bb + g|\Psi_0|^2),\ \xi_1 = 2\mathrm{Re}(g\Psi_0^2),\ \xi_2 = -2\mathrm{Im}(g\Psi_0^2)$ and $\Psi_0 = \sqrt{N_0/V}e^{i\theta}$ is the condensate wavefunction.  The operators in Eq.(\ref{ham}) are defined as
\begin{eqnarray} \label{generators}
  K_0 &=& \frac{1}{2}(\ak^\dag \ak + \amk\amk^\dag), \nn\\
  K_1 &=& \frac{1}{2}(\ak^\dag \amk^\dag + \ak\amk), \nn\\
  K_2 &=& \frac{1}{2i}(\ak^\dag\amk^\dag - \ak\amk),
\end{eqnarray}
They are the three generators of $\mathfrak{su}(1,1)$ Lie algebra \cite{lyu_geometrizing_2020, puri_mathematical_2001}, which satisfy the $\mathfrak{su}(1,1)$ Lie algebra commutation relation
\begin{eqnarray} \label{commu}
  [K_0, K_1] = iK_2,\, [K_2, K_0] = iK_1, [K_1, K_2] = -iK_0.
\end{eqnarray}
Since the Bogoliubov Hamiltonian $\hh_{\mathrm{Bog.}}$ is decoupled for different momenta, in the following we will focus on the Hamiltonian $\hh_\bb$ in Eq.(\ref{ham}). This Hamiltonian can be diagonalized by the Bogoliubov transformation, which introduces the quasi-particle excitations as follows
\begin{eqnarray}
\hat{\alpha}_\bk = \uk a_\bk + \vk a_{-\bk}^\dag, \ \
\hat{\alpha}_{-\bk} = \uk a_{-\bk} + v_\bk a_\bk^\dag,
\label{eq-bogo}
\end{eqnarray}
where the coefficients $\uk, \vk$ satisfy the condition
\begin{eqnarray} \label{sp}
  |\uk|^2 - |\vk|^2 = 1.
\end{eqnarray}
In terms of quasi-particle operators, the Hamiltonian $\hh_\bb$ can be diagonalized as $\hh_\bb=\varepsilon_\bb(\hat{\alpha}_\bk^\dag\hat{\alpha}_\bk+\hat{\alpha}_{-\bk}^\dag\hat{\alpha}_{-\bk}) + \mathrm{const}.$ with the quasi-particle energy $\varepsilon_\bb = \sqrt{\xi_0(\bk)^2 - \xi_1(\bk)^2 - \xi_2(\bk)^2}/2$.

In order to compute the Uhlmann phase later, we also need to work out the explicit form of all eigenstates. The ground state of the Hamiltonian $\hh_\bk$ is defined by $\ha |G\rangle = \hma |G\rangle = 0$. One can verify that the ground state is just the generalized coherent state of $SU(1,1)$ group \cite{perelomov_generalized_1986, zhai_ultracold_2021}
\begin{eqnarray}
|G\rangle = \frac{1}{\sqrt{1 - |z|^2}} e^{-z a_\bk^\dag a_{-\bk}^\dag} |0\rangle,
\end{eqnarray}
where $z = \vk / \uk$, and $|0\rangle$ is the bare vacuum state. Due to Eq.(\ref{sp}), the norm of $z$ is smaller than 1, i.e. $1 - |z|^2 > 0$. Therefore, for all the possible $\hh_\bk$, their ground states are parameterized by $z$ defined on a two-dimensional unit disk, which is actually the celebrated Poincare disk \cite{lyu_geometrizing_2020, perelomov_generalized_1986}.

On the other hand, the ground state $|G\rangle$ can also be written as a unitary operator $D(\zeta)$ acting on the vacuum state \cite{perelomov_generalized_1986}
\begin{eqnarray}\label{d_op}
  |G\rangle = D(\zeta) |0\rangle \equiv e^{-\zeta a_{\bk}^\dag a_{-\bk}^\dag + \zeta^* a_{\bk}a_{-\bk}} |0\rangle,
\end{eqnarray}
which will be useful for the calculation of Uhlmann phase. Meanwhile, if we write a point $z$ in the Poincare disk as $z = \tanh\frac{r}{2}e^{i\theta}$ with $r\in [0,\infty), \ \theta\in [-\pi, \pi)$, then $\zeta$ is given as $\zeta = \frac{r}{2}e^{i\theta}$ \cite{hasebe_sp_2020}. With this parameterization of $\zeta$, we can also make the Euler decomposition of the operator $D(\zeta)$ as follows \cite{hasebe_sp_2020}
\begin{eqnarray} \label{euler}
  D(\zeta) = e^{i\theta K_0} e^{-i r K_2} e^{-i \theta K_0},
\end{eqnarray}
which is also needed when calculating the Uhlmann phase.

With the ground state and the creation operators of quasi-particles, we can obtain the excited state
\begin{eqnarray}\label{psi_n}
  |\Psi_\mathbf{n}\rangle = \frac{1}{\sqrt{n_1!n_2!}} (\ha^\dag)^{n_1}(\hma^\dag)^{n_2} |G\rangle,
\end{eqnarray}
where $\mathbf{n} = (n_1,n_2)$ labels the number of each kind of quasi-particles. Note that the Bogoliubov transformation of Eq.(\ref{eq-bogo}) can also be written as \cite{puri_mathematical_2001, perelomov_generalized_1986}
\be
D(\zeta)^\dag \hat{\alpha}_{\pm\bb}^\dag D(\zeta) = a_{\pm\bb}^\dag\nonumber
\ee 
With this fact, it can be shown that
\begin{eqnarray}\label{dn}
|\Psi_\mathbf{n}\rangle = \frac{1}{\sqrt{n_1!n_2!}} D(\zeta) (\ak^\dag)^{n_1}(\amk^\dag)^{n_2} |0\rangle=D(\zeta) |\mathbf{n}\rangle,
\end{eqnarray}
where we have defined 
\be
\ket{\bf{n}} = \frac{1}{\sqrt{n_1!n_2!}}(\ak^\dag)^{n_1}(\amk^\dag)^{n_2} |0\rangle.
\ee 
Thus, one can see that the excited states $|\Psi_\bf{n}\rangle$ are also parameterized by a point $z$ on the Poincare disk. And the eigenenergy of this excited state is given by $E_{\mathbf{n}}=\varepsilon_\bk(n_1+n_2)$. Here we have shifted the ground state energy to 0 for simplicity. This shift will not affect the calculation of Uhlmann phase in the following sections.

\section{Uhlmann process and Uhlmann connection} \label{sec:uhl}

The Uhlmann phase is a generalization of the celebrated Berry phase for quantum systems at finite temperature and mixed quantum states \cite{uhlmann_parallel_1986, uhlmann_gauge_1991, uhlmann_density_1993}. In this section, we will briefly review the concept of Uhlmann connection and Uhlmann phase. 

\begin{figure}
  \includegraphics[width = .45\textwidth]{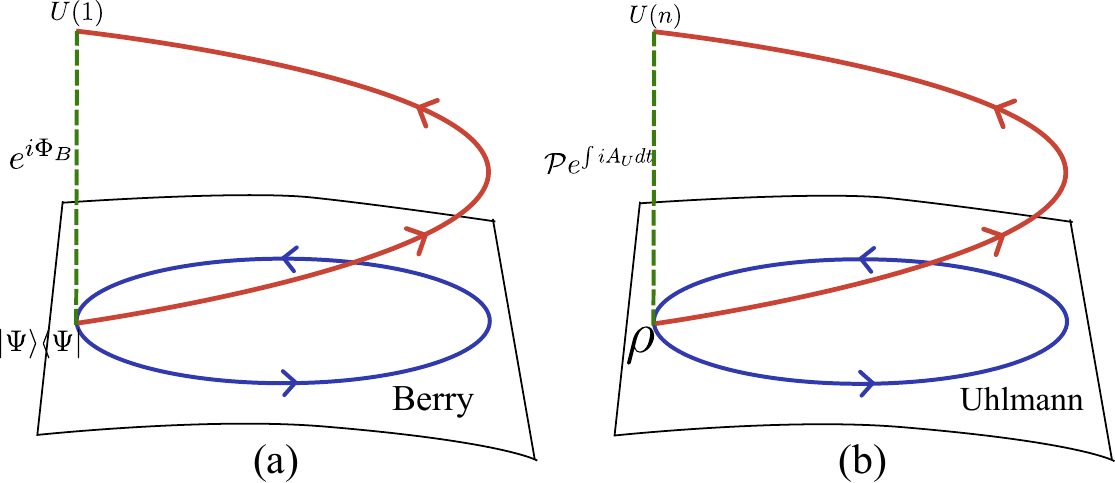}
  \caption{The comparison for the definitions of Berry phase and Uhlmann phase.}
  \label{fig:ber_uhl}
\end{figure}

When a quantum system undergoes an adiabatic cyclic process in the parameter space, its $n$-th eigenstate $\ket{\psi_n(t)}$ with $t\in [0,1]$ parameterizing the adiabatic path on the parameter space, will acquire a phase factor, i.e. $\ket{\psi_n(1)} = e^{i\Phi_B}\ket{\psi_n(0)}$, where $\Phi_B$ is the Berry phase. The Berry phase can be obtained by integrating the Berry connection $A_B=-i\ep{\psi_n|\frac{d}{d t}|\psi_n}$, i.e. $\Phi_B = \int A_B d t$, as illustrated in Figure \ref{fig:ber_uhl}(a).

For a finite-temperature system or a mixed state, its physical properties can be computed from the density matrix $\rho$ instead of wave functions. One can always make a spectral decomposition of density matrices as $\rho=\sum_i \lambda_i\ket{u_i}\bra{u_i}$. Here $\ket{u_i}$ for $i=1,\cdots,n$ are eigenstates and $\lambda_i=e^{-E_i/T}/\mathcal{Z}$ is the Boltzmann weight of eigenstates $\ket{u_i}$ for systems at thermal equilibrium, where $\mathcal{Z} = \sum_i e^{-E_i/T}$. Here, for simplicity, we have set the Boltzmann constant $k_B$ as 1.

In parallel to the Berry phase, for a finite-temperature system undergoing a cyclic adiabatic process, with density matrix denoting as $\rho(t)$ and $t\in[0,1]$ representing the adiabatic parameter, Uhlmann phase can be defined as the argument of the trace of the density matrix multiplied by an $SU(n)$ unitary transformation $W$ which plays a similar role as the $U(1)$ factor in the Berry phase case, i.e.
\begin{eqnarray}\label{uhl_ph_def}
  \Phi_U = \mathrm{Arg}\tr[\rho(0)W],
\end{eqnarray}
as illustrated in Figure \ref{fig:ber_uhl}(b). Here $W$ is the so-called Wilson loop defined as the integration of the Uhlmann connection $A_U$, i.e. 
\begin{eqnarray} \label{wilson}
  W=\mathcal{P}\exp\Big(\int A_U d t\Big),
\end{eqnarray}
where $\mathcal{P}$ denotes the path ordering. The Uhlmann connection $A_U$ is a generalization of Berry connection to an $SU(n)$ gauge connection for mixed states, and is given by 
\be\label{uhl_con_0}
A_U=\sum_{m\neq n}\frac{(\sqrt{\lambda_m}-\sqrt{\lambda_n})^2}{\lambda_m+\lambda_n}\ket{u_m}\bra{u_m}\frac{d}{d t} \ket{u_n}\bra{u_n}.
\label{eq-AU}
\ee
For later convenience, we also define the Uhlmann overlap as 
\begin{eqnarray}
  \mathcal{F} = \tr[\rho(0)W].
\end{eqnarray} 
The Uhlmann has already been used to study several topological systems at finite temperatures \cite{viyuela_twodimensional_2014}. In the rest of this section, we will give a brief derivation of the Uhlmann connection. And more details can be found in \cite{viyuela_symmetryprotected_2015}.

The concept of Uhlmann connection is based on decomposition of the density matrix as
\be
\rho=w w^{\dagger},\qquad w=\sqrt{\rho}\,U.
\ee
Here $w$ can be thought of as the counterpart of the wave function for the mixed states. For a given $\rho$, the amplitude $w$ contains an arbitrary $U(n)$ phase factor. If we make the replacement $w\to wU$ where $U$ arbitrary is unitary matrix, $\rho$ will stay the same.

Although the overall phase factor is only a gauge choice, we can determine the relative phase factor $U$ by the following parallel condition proposed by Uhlmann \cite{uhlmann_parallel_1986}:
\be\label{para}
w_1^{\dagger}w_2=w_2^{\dagger}w_1=C>0.
\ee
Here $C>0$ means that $C$ is a Hermitian and positive definite matrix. With this condition, the relative phase factor between $w_1$ and $w_2$ is uniquely determined as long as the density matrices are always full rank. Consider two different density matrices and their amplitudes $w_1=\sqrt{\rho_1}U_1$ and $w_2=\sqrt{\rho_2}U_2$. The parallel condition determines that
\be
U_2U_1^{\dagger}=\sqrt{\rho_2^{-1}}\sqrt{\rho_1^{-1}}\sqrt{\sqrt{\rho_1}\rho_2\sqrt{\rho_1}}.
\label{eq-UU}
\ee
The above result is a finite version of the Uhlmann connection. Here $\rho$ is assumed to be a full-rank matrix, since the inverse of $\rho$ has been used.

For the adiabatic path and the associated density matrix $\rho(t)$, we seek a decomposition of the form $\rho(t) = w(t)w^\dag(t)$ with $w(t) = \sqrt{\rho}W(t)$, such that adjacent amplitudes $w(t)$ and $w(t+dt)$ satisfy the parallel condition in Eq.(\ref{para}), namely $w^\dag(t)w(t+dt) = w^\dag(t+dt)w(t) > 0$. The Uhlmann connection is defined as the differential generator of $W(t)$, expressed as $W(t) = \mathcal{P}e^{\int_0^t A_U(t')dt'}$, where $\mathcal{P}$ denotes path ordering. Taking the derivative of $W(t)$ yields $dW(t)/dt = A_U(t)W(t)$, which can be rearranged as
\begin{eqnarray}
A_U(t) = \frac{d W(t)}{d t} W^\dagger(t).
\end{eqnarray}

Then by substituting $\rho(t) = \sum_i \lambda_i(t) \ket{u_i(t)}\bra{u_i(t)}$ to above expression,  and utilizing Eq.(\ref{eq-UU}), which is derived from the parallel condition Eq.(\ref{para}), one can obtain the Uhlmann connection Eq.(\ref{uhl_con_0}). The detailed steps of this derivation are provided in Appendix \ref{app:detail} for readers interested in the full mathematical process. One can verify that $A_U$ will transform like an ordinary $U(n)$ non-Abelian gauge field under a gauge transformation.

\section{The Uhlmann phase of BEC}\label{sec:uhl_bec}

With the definitions of Uhlmann connection and Uhlmann phase in hand, in this section, we will study the Uhlmann phase of BEC at finite temperature. 

By adiabatically tuning the parameters in Hamiltonian Eq.(\ref{ham}), denoted as $\hh_\bb(t)$ with $t \in [0,1]$ representing the parameter, the corresponding density operator is given by $\rho(t) = \sum_\bf{n} \lambda_\bf{n}(t) \ket{\Psi_\bf{n}(t)}\bra{\Psi_\bf{n}(t)}$ in temperature $T$. Here, $\lambda_\bf{n}(t) = e^{-E_\bf{n}(t)/T}/\mathcal{Z}(t)$ where $\mathcal{Z}(t) = \sum_\mathbf{n}e^{-E_\bf{n}(t)}/T$, and the states $\ket{\Psi_\bf{n}(t)}$ are defined in Eq.(\ref{psi_n}). By noting Eq.(\ref{dn}), one can see that the density operator $\rho(t)$ at each $t$ also corresponds to a point $z(t)$ in the Poincare disk, which can be parameterized as 
\begin{eqnarray}\label{zt}
  z(t) = \tanh\frac{r(t)}{2}e^{i\theta(t)}.
\end{eqnarray} 
This correspondence gives us a geometric way to visualize the Uhlmann process for the BEC, i.e. the density operators $\rho(t)$ with $t\in[0,1]$ corresponds to a path in the Poincare disk, as shown in Figure \ref{paths}. 

Thus, by substituting the expression Eq.(\ref{dn}) for the states $|\Psi_\bf{n}\rangle$ into the definition of Uhlmann connection Eq.(\ref{uhl_con_0}), and making use of the properties of operator $D(\zeta)$ and its Euler decomposition Eq.(\ref{euler}), after some straightforward but tedious calculations (see appendix \ref{app_uhl} for more details), one can obtain the Uhlmann connection of BEC
\begin{eqnarray}\label{a_u}
  A_U &=& -i\eta \Big[\dot{\theta} \sinh^2r K_0 + (\dot{r}\sin\theta + \frac{1}{2}\dot{\theta}\sinh(2r) \cos\theta)K_1 \nn\\
  &&+ (\dot{r}\cos\theta - \frac{1}{2}\dot{\theta}\sinh(2r) \sin\theta)K_2 \Big],
\end{eqnarray}
where $\eta = 1 - 1/\cosh(\beta\varepsilon_\bb)$, and $r(t),\ \theta(t)$ representing the path in the Poincare disk as defined in Eq.(\ref{zt}). Then we need to calculate the Uhlmann overlap
\begin{eqnarray}
  \mathcal{F}(T) &=& \tr[\rho(0)W] \nn\\
  &=& \sum_{\mathbf{n}} \lambda_\mathbf{n}(0) \langle \mathbf{n}|D^\dag(\zeta_0) W D({\zeta_0}) |\mathbf{n}\rangle \nn\\
  &\equiv& \sum_{\mathbf{n}} \lambda_\mathbf{n}(0)\langle \mathbf{n}| U |\mathbf{n}\rangle,
\end{eqnarray}
where we have defined $U = D^\dag(\zeta_0) W D({\zeta_0})$, and $\lambda_\mathbf{n}(0) = e^{-E_\mathbf{n}(0)/T}/\mathcal{Z}$. We note that the Uhlmann connection Eq.(\ref{a_u}) is a linear combination of the $\mathfrak{su}(1,1)$ generators, thus, the Uhlmann Wilson operator $W$ corresponds to an element in the $SU(1,1)$ group, which is similar to the time evolution operator of the BEC. Meanwhile, the operator $D(\zeta_0)$ is also an exponential of a linear combination of the $\frak{su}(1, 1)$ Lie algebra generators, and thus corresponds to an element in the $SU(1,1)$ group. Consequently, the operator $U = D^\dag(\zeta_0) W D({\zeta_0})$ is a product of three $SU(1,1)$ group elements, which itself corresponds to an element in the $SU(1,1)$ group. Since every element of the $SU(1,1)$ group has the so-called normal order decomposition and the operator $U$ corresponds to an element in this $SU(1,1)$ group, thus $U$ has normal order decomposition as \cite{perelomov_generalized_1986, puri_mathematical_2001}
\begin{eqnarray} \label{gauss_decomp}
  U = e^{\mu K^+}e^{\nu K_0}e^{\mu' K^-},
\end{eqnarray}
where $K^+ = a_{\bk}^\dag a_{-\bk}^\dag, K^- = a_{\bk} a_{-\bk}$. And the factors $\mu, \mu'$ and $\nu$ can be calculated numerically, which we will discuss soon. Then, by doing Taylor expansion on the exponential terms of Eq.(\ref{gauss_decomp}), and keeping the non-zero terms, we have
\begin{eqnarray}
  &&\langle \mathbf{n}|U|\mathbf{n}\rangle \nn\\
  &=& \sum_{m=0}^{\mathrm{min}\{n_1,n_2\}} \frac{e^{\frac{\nu}{2}(n + 1)}(e^{-\nu} \mu \mu')^m}{(m!)^2} \frac{n_1!n_2!}{(n_1-m)!(n_2-m)!} \nn\\
   &=& e^{\frac{\nu}{2}(n + 1)} {}_2F_1(-n_1, -n_2, 1, e^{-\nu} \mu\mu'),
\end{eqnarray}
where $_2F_1$ is the hypergeometric function and $n = n_1 + n_2$. Hence, we have the expression for the Uhlmann phase 
\begin{eqnarray}\label{eq:phi_u}
  \Phi_U = \mathrm{Arg}\sum_{\mathbf{n}} \lambda_\bf{n}(0) e^{\frac{\nu}{2}(n+1)} {}_2F_1(-n_1, -n_2, 1, e^{-\nu} \mu\mu'). \hspace{0.2cm}
\end{eqnarray}
Noting that $A_U/i$ is Hermitian, thus, $U$ is a unitary operator, which indicates $|\langle \bf{n}|U| \bf{n}\rangle| < 1$. Hence, in above expression, the summation converges and $\Phi_U$ can be calculated numerically by choosing suitable cutoffs for $n_1, n_2$.

\subsection{Analytical result for circle paths}
To compute the Uhlmann phase, the difficult part is the Uhlmann Wilson loop Eq.(\ref{wilson}) which is a path ordering product $W=\mathcal{P}\exp\Big(\int A_U d t\Big)$. Since $A_U$ with different parameters usually do not commute, the path ordering product is quite non-trivial and has to be computed numerically in general.

\begin{figure}
  \includegraphics[width = .4\textwidth]{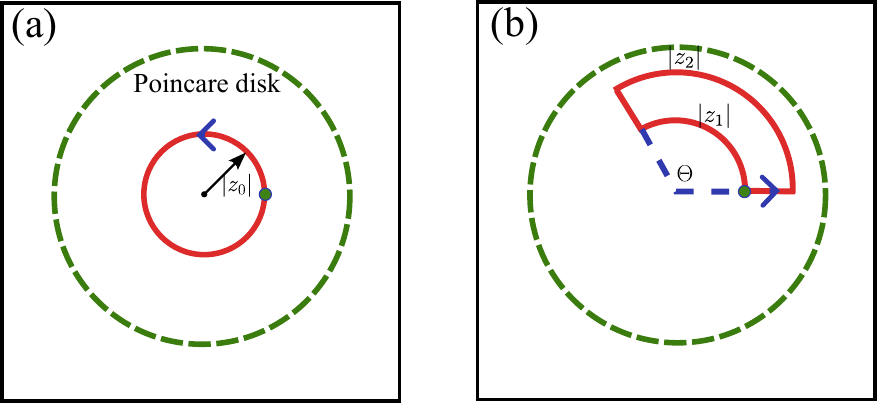}
  \caption{The schematics for the paths of adiabatic Uhlmann process in the Poincare disk, where the red solid lines are the paths, and the green dots are initial points, and the blue arrows indicate the direction of the adiabatic process. (a) Circle path with center coincide with the center of the Poincare disk and radius $|z_0| = \tanh(r_0/2)$. (b) The inner arc has radius $|z_1| = \tanh(r_1/2)$ and the outer arc has radius $|z_2| = \tanh(r_2/2)$.}
  \label{paths}
\end{figure}

However, for the circle path whose center coincides with the center of the Poincare disk as shown in Figure \ref{paths}(a), we can analytically compute $W$. For this kind of path, $r$ is constant, thus, the $\dot{r}$ terms in Eq.(\ref{a_u}) vanish. In this case, we define the Wilson line along an arc as
\be
W(\theta)=\mathcal{P}\exp\Big(\int_0^{\theta} A_U(r,\theta') d\theta'\Big),
\ee
where we have converted the integration over the parameter $t$ to one over $\theta'$. It satisfies the following differential equation
\be
\frac{d W(\theta)}{d\theta}&=&A_U W(\theta), \nn\\
A_U&=&-U_1\Big[K_1\cosh r+K_0\sinh r\Big]U_1^\dag\cdot(i\eta\sinh r). \hspace{0.8cm}
\ee 
Here $U_1=e^{i\theta K_0}$. If we set $W=U_1\widetilde{W}$, the above equation is simplified to
\be
\frac{d\widetilde{W}}{d\theta}&=&\Big[U_1^\dag A_U U_1-U_1^\dag\frac{d U_1}{d\theta}\Big]\widetilde{W}\nonumber\\
&=&\Big[-i\eta\sinh r(K_1\cosh r+K_0\sinh r)-i K_0\Big]\widetilde{W}. \hspace{0.8cm}
\ee
Now the terms inside the bracket do not depend on $\theta$. It is easy to solve the above equation to find that
\be
W(2\pi)=e^{2\pi i K_0}e^{-2\pi i\Big[K_0+\eta\sinh r(K_1\cosh r+K_0\sinh r)\Big]}.
\ee

\subsection{General numerical method for calculating Uhlmann phase of BEC}
For a general path, it is difficult to calculate the Uhlmann Wilson loop operator analytically. Thus, we resort to the help of the representation theory of $SU(1,1)$ group, i.e. the Uhlmann Wilson loop operator uniquely corresponds to a matrix of the $SU(1,1)$ group.

This correspondence can be seen by noticing that the Uhlmann connection Eq.(\ref{a_u}) is a linear combination of the $\mathfrak{su}(1,1)$ Lie algebra. On the other hand, the $\mathfrak{su}(1,1)$ Lie algebra also has a matrix representation, i.e. the operators $K_{0,1,2}$ corresponds to the matrices
\begin{eqnarray}
  K_0 \leftrightarrow \sigma_z/2,\ K_1 \leftrightarrow i\sigma_y/2,\ K_2 \leftrightarrow -i\sigma_x/2,
\end{eqnarray}
where $\sigma_{x,y,z}$ are the three Pauli matrices. One can verify that the matrices $\{\sigma_z/2, i\sigma_y/2, -i\sigma_x/2\}$ also satisfy the commutation relation Eq.(\ref{commu}) of $\mathfrak{su}(1,1)$ Lie algebra. By replacing the operators $K_{0,1,2}$ with their corresponding matrices, one can obtain the matrix representation of the Uhlmann connection $\mathcal{A}_U$. Then, one have the corresponding matrix $\mathcal{W} = \mathcal{P}e^{\int \mathcal{A}_U d t}$, which belong to the $SU(1,1)$ group, of the Uhlmann Wilson loop operator. We note that the matrix can be calculated numerically by slicing the path into \st{to} $N$ pieces, i.e.
\begin{eqnarray}
  \mathcal{W} = \prod_{i=1}^N e^{\mathcal{A}_U(t_i)\Delta t}.
\end{eqnarray}
Similarly, one can obtain the matrices corresponding to the operators $K^{\pm}$, i.e. $K^+ \leftrightarrow \sigma^+,\ K^- \leftrightarrow -\sigma^-$, where $\sigma^{\pm} = (\sigma_x \pm i\sigma_y)/2$. Thus, one can calculate the corresponding matrix of operator $D(\zeta_0)$ defined in Eq.(\ref{d_op}), i.e. $\mathcal{D}(\zeta_0) = e^{-\zeta_0 \sigma^+ - \zeta_0^* \sigma^-}$. As a result, the normal order decomposition of Eq.(\ref{gauss_decomp}) can be cast on the matrix representation level, which is just the Gaussian decomposition of the matrix $\mathcal{U} = \mathcal{D}^\dag(\zeta_0)\mathcal{W}\mathcal{D}(\zeta_0)$, i.e. Eq.(\ref{gauss_decomp}) corresponds to 
\begin{eqnarray}
  \mathcal{U} = \left(
    \begin{array}{cc} 1 & \mu \\ 0 & 1\end{array}
  \right)
  \left(
    \begin{array}{cc} e^{\frac{\nu}{2}} & 0 \\ 0 & e^{-\frac{\nu}{2}}\end{array}
  \right)
  \left(
    \begin{array}{cc} 1 & 0 \\ -\mu' & 1\end{array}
  \right).
\end{eqnarray}
Hence, one can readily obtain the factors $\mu, \mu'$ and $\nu$ in terms of the matrix elements of $\mathcal{U}$, which can be calculated numerically in general \cite{perelomov_generalized_1986}
\begin{eqnarray}
  \mu = \frac{\mathcal{U}_{12}}{\mathcal{U}_{22}},\ \mu' = -\frac{\mathcal{U}_{21}}{\mathcal{U}_{22}},\ \nu = -2\ln\mathcal{U}_{22}.
\end{eqnarray}
Then one can substitute these factors to Eq.(\ref{eq:phi_u}), and take proper cutoff for the energy level for $n_1$ and $n_2$ to calculate the Uhlmann phase.

\begin{figure}
  \includegraphics[width = .45\textwidth]{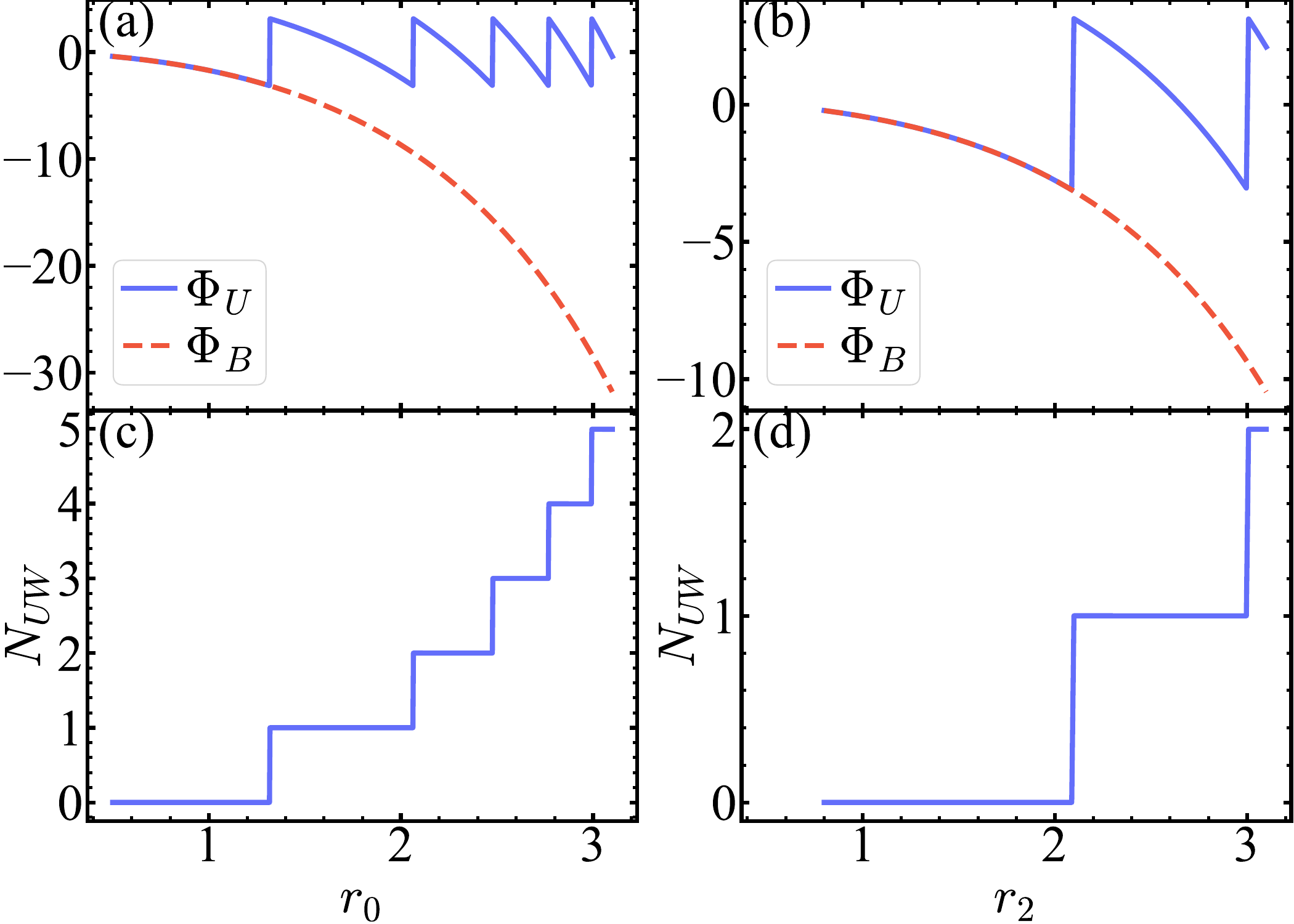}
  \caption{Comparison between the Uhlmann phase in low temperature limit ($T=0.01$ here) and the Berry phase at zero temperature. (a) The adiabatic path is the one in Figure \ref{paths}(a) with varying $r_0$. (b) The adiabatic path is the one in Figure \ref{paths}(b) with $r_1=1/2$ and varying $r_2$. (c) and (d) show the values of $N_{UW}$ for the cases in (a) and (b), respectively.}
  \label{uhl_berry}
\end{figure}

\section{The Uhlmann phase winding of BEC} \label{sec:uhl_wind}

Having discussed the general method of calculating the Uhlmann phase of BEC, in this section, we will present the numerical results of the Uhlmann phase for different paths on the Poincare disk. These results reveal the unique feature of the Uhlmann phase in the BEC system. Without loss of generality, we assume that the eigenenergy of the quasi-particles is fixed \st{to $\varepsilon_\bk=1$} in our numerical calculations. It also serves as an energy unit in the following numerical results.

We first consider the adiabatic process where the path is a circle on the Poincare disk with the center located at the origin of the Poincare disk as shown in Figure \ref{paths}(a). The parameter equations of this path are
\begin{eqnarray}\label{circle}
  r(t) = r_0, \ \theta(t) = 2\pi t,
\end{eqnarray}
where $r_0$ is a constant, and $t\in [0,1]$. Since the Uhlmann phase is a finite temperature generalization of Berry phase, one may expect that the Uhlmann phase would approach the Berry phase in the zero temperature limit. This is true for previous studies of Uhlmann phase of topological models \cite{viyuela_twodimensional_2014} or spin systems \cite{hou_finitetemperature_2021}. However, this is not exactly the case for the BEC. The Berry phase of the $SU(1,1)$ coherent states for the adiabatic process with path described by Eq.(\ref{circle}) is \cite{damaskinski_calculation_1991}
\begin{eqnarray}
  \Phi_B = -2\pi\sinh^2\frac{r_0}{2}.
\end{eqnarray}
Note that the above Berry phase takes values in negative real numbers. The comparison between the Uhlmann phase and Berry phase is shown in Figure \ref{uhl_berry}(a). One can see that the Uhlmann phase coincide with Berry phase only when $r_0$ is small. The reason is that the Uhlmann phase by definition of Eq.(\ref{uhl_ph_def}) is the phase angle of the Uhlmann overlap $\mathcal{F}(T)$, which indicating that $\Phi_U\in [-\pi, \pi)$. Thus, for the BEC case, the Uhlmann phase can be different from the Berry phase by some multiples of $2\pi$. More explicitly, we have
\begin{eqnarray}
\lim_{T\to0}\Phi_U=\Phi_B + 2\pi N_{UW}.
\label{eq-U0}  
\end{eqnarray}
In Fig. \ref{uhl_berry}(c), we show the values of $N_{UW}$ varying with the radius $r_0$ of a circle path. However, one may wonder whether the $2\pi$-multiple makes any difference, since the Berry phase is just the phase factor acquired by the ground state in the adiabatic evolution $\ket{G(1)} = e^{i\Phi_B}\ket{G(0)}$. Actually, as we will see later, this $2\pi$-multiple does have its meaning, and can have measurable effect once we consider the behavior of the Uhlmann overlap in the finite temperature range.

\begin{figure}
  \includegraphics[width = .48\textwidth]{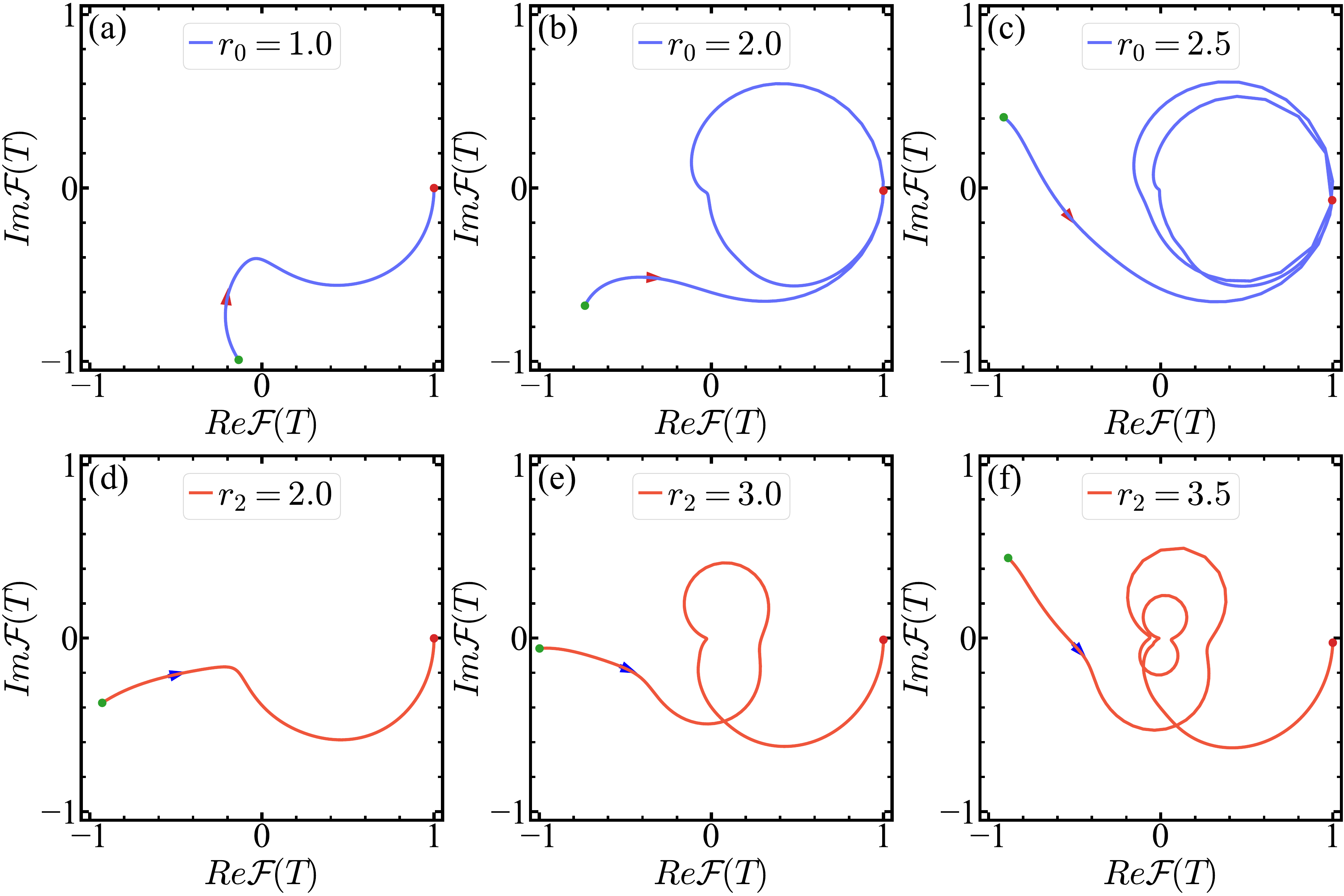}
  \caption{$\mathcal{F}(T)$ from $T/\varepsilon_\bk = 0.01$ to $T/\varepsilon_\bk = 500$. The direction of the arrow indicates that the temperature is from low to high. The cutoff of energy level is taken as $n_1 = n_2 = 50$. (a)-(c) The adiabatic path is the one in Figure \ref{paths}(a). (d)-(f) The adiabatic path is the one in Figure \ref{paths}(b) with $r_1 = 0.5$.}
  \label{fig_uhl}
\end{figure}

\begin{figure}
  \includegraphics[width = .45\textwidth]{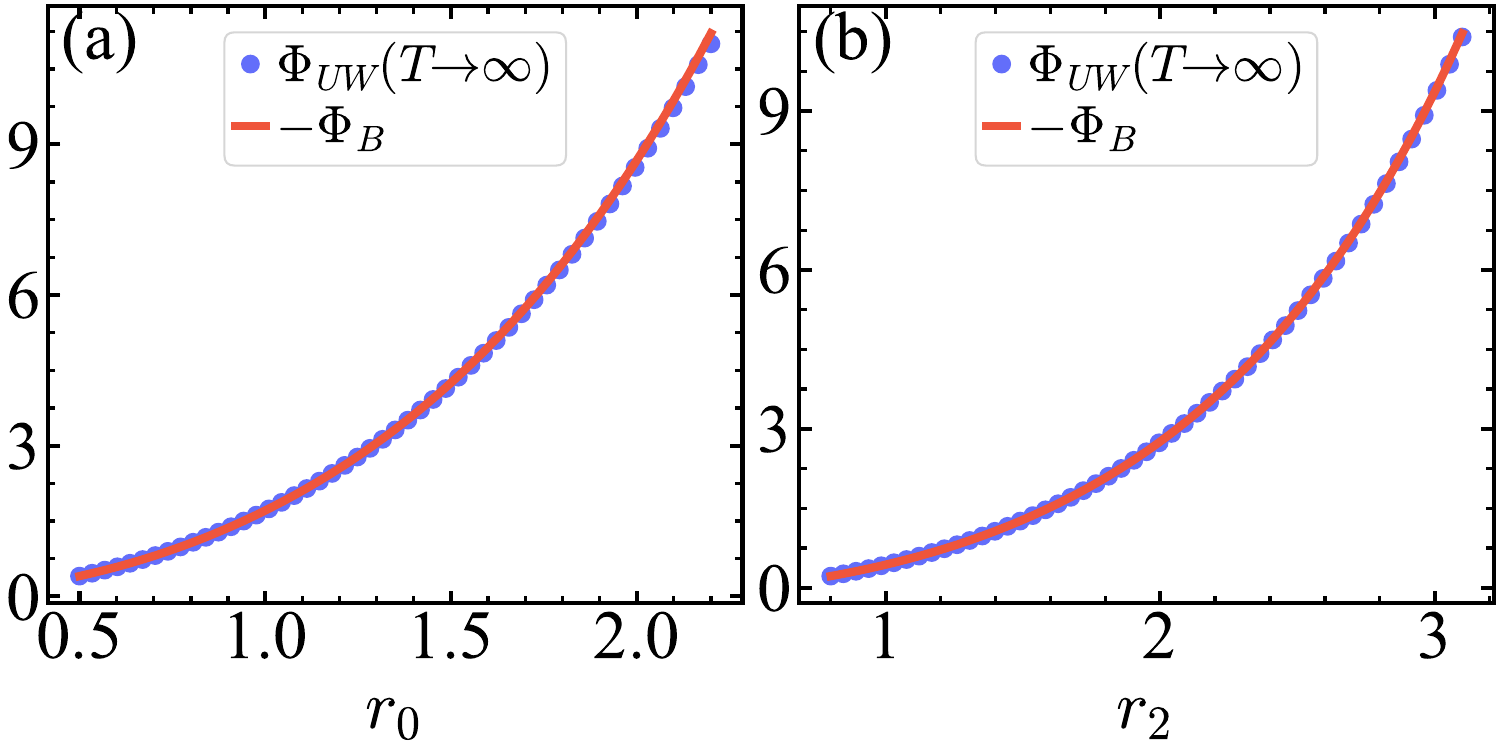}
  \caption{The comparison between the Uhlmann phase winding $\Phi_{UW}$ and the negative of Berry phase $-\Phi_B$ for (a) the path shown in Figure \ref{paths}(a) and (b) the path shown in Figure \ref{paths}(b).}
  \label{fig_uhl_wind}
\end{figure}

\begin{figure}
  \includegraphics[width = .2\textwidth]{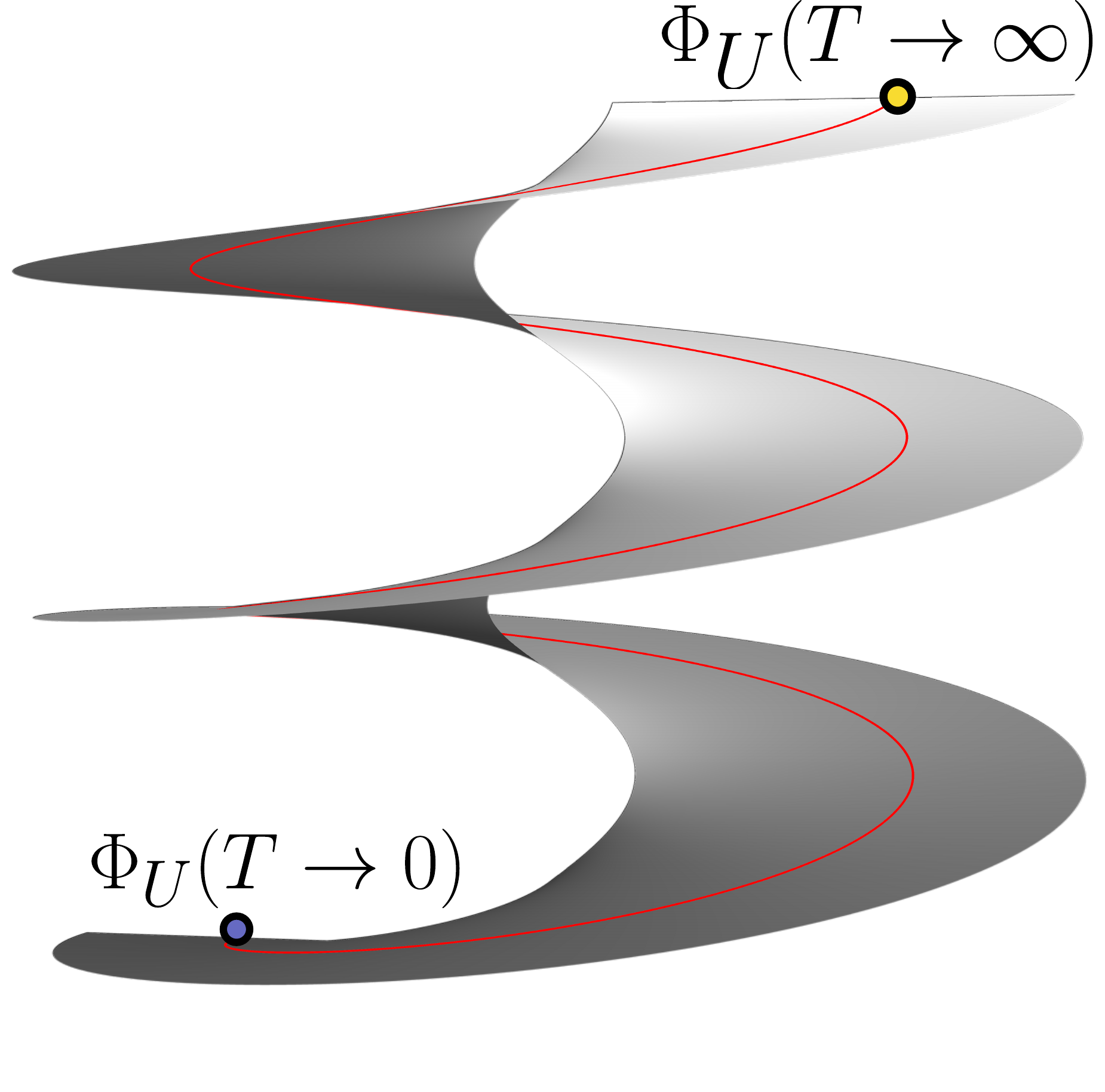}
  \caption{Schematics for the Uhlmann phase takes values in a Riemann surface as the temperature increases from zero temperature limit to infinite temperature limit.}
  \label{riemann}
\end{figure}

In order to provide a better visualization of how the Uhlmann phases change with temperature, in Figure \ref{fig_uhl}(a)-(c), we plot the trajectories of the Uhlmann fidelity $\mathcal{F}(T)$ on the complex plane as the temperature varies from very low as $T/\varepsilon_\bk=0.01$ to very high as $T/\varepsilon_\bk=500$. The adiabatic paths are still circles on the Poincare disk as in Figure \ref{paths} (a). From these figures, we can see that as the circle radius $r_0$ increases, the Uhlmann overlap starts to circle around the origin of the complex plane as the temperature increases. And as $T\to \infty$, the Uhlmann phase approach to 0, which is the same as previous studies \cite{viyuela_twodimensional_2014}.

To have a better understanding of the winding behavior of the Uhlmann overlap $\mathcal{F}(T)$, we can study the total degrees that the Uhlmann phase winds around the origin as the temperature increases from zero to infinity. This consideration suggests the following definition of Uhlmann phase winding
\be
\Phi_{UW}(T) = \int_0^T \p_T'\Phi_U(T') d T'.
\ee
In practice, we computed the Uhlmann phase winding numerically by integrating from $T=0.01$ to $T=500$. In Figure \ref{fig_uhl_wind} (a), we plot the Uhlmann phase winding as a function of the path radius $r_0$. One can see that the Uhlmann phase winding $\Phi_{UW}$ always perfectly agrees with $-\Phi_B$. From this numerical result, one can deduce that
\begin{eqnarray}
\Phi_{UW}(T\to \infty) = -\Phi_B,
\label{eq-UWB}
\end{eqnarray}
In other words, the total degree that the Uhlmann phase winds from zero to infinite temperature is equal to negative of the Berry phase at zero temperature. This result can also be understood as follows. Since Uhlmann phase is always zero at infinite $T$, we find that
\be
\Phi_{UW}(T\to \infty)=\lim_{T\to\infty}\Phi_U-\lim_{T\to 0}\Phi_U+2\pi N_{UW}=-\Phi_B, \hspace{.2cm}
\ee
where Eq.(\ref{eq-U0}) has been used. Thus, the $2\pi$ multiple in the Berry phase is related to the Uhlmann phase winding. And as we propose a method to observe the Uhlmann phase, hence, this $2\pi$ multiple in Berry phase can have measurable effect.

One may wonder whether the above result of Eq.(\ref{eq-UWB}) is only valid for the circular symmetric paths. To answer this question, we also calculate the Uhlmann phase for adiabatic paths that enclose a sector of the Poincare disk as shown in Figure \ref{paths}(b), with the inner radius $|z_1| = \tanh(r_1/2)$ and outer radius $|z_2| = \tanh(r_2/2)$. In our numerical calculations, we always fix the arc angle as $\Theta = 2\pi/3$ and the inner radius as $r_1 = 1/2$. Then the Uhlmann phase in the zero temperature limit is plotted as a function of outer radius $r_2$ in comparison the Berry phase in Figure \ref{uhl_berry}(b). Again, one can see that they are different by some multiple of $2\pi$. 

In Figure \ref{fig_uhl}(d)-(f), the curves of the Uhlmann overlap $\mathcal{F}(T)$ on the complex plane display very similar behaviors as before. Therefore, we can still compute the Uhlmann phase winding for this type of paths and the resulting $\Phi_{UW}$ also agrees with $-\Phi_B$ as shown in Figure \ref{fig_uhl_wind}(b). These results of sector-like paths suggest that the relation of Eq.(\ref{eq-UWB}) between the Uhlmann phase winding and Berry phase is not limited to the circular paths but a general feature of the BEC.

Thus, we conclude that the Uhlmann phase, as a function of temperature, winds from $\Phi(T \to 0)$ to $0$ as the temperature increases from the zero-temperature limit to the infinite-temperature limit. To further reveal the geometric meaning of the Uhlmann phase winding, it is suggested that the Uhlmann phase takes values in a Riemann surface. Actually, we can view the Uhlmann phase as a multiple valued function $\ln(\mathcal{F}/|\mathcal{F}|)/i$. It can be converted into a single valued function which is defined on a Riemann surface of $\ln(z)$ as shown in Figure \ref{riemann}. Then the geometric meaning of $N_{UW}$ is just the number of times that the Uhlmann phase circles around the branch point on the Riemann surface.

\begin{figure}
  \includegraphics[width = .45\textwidth]{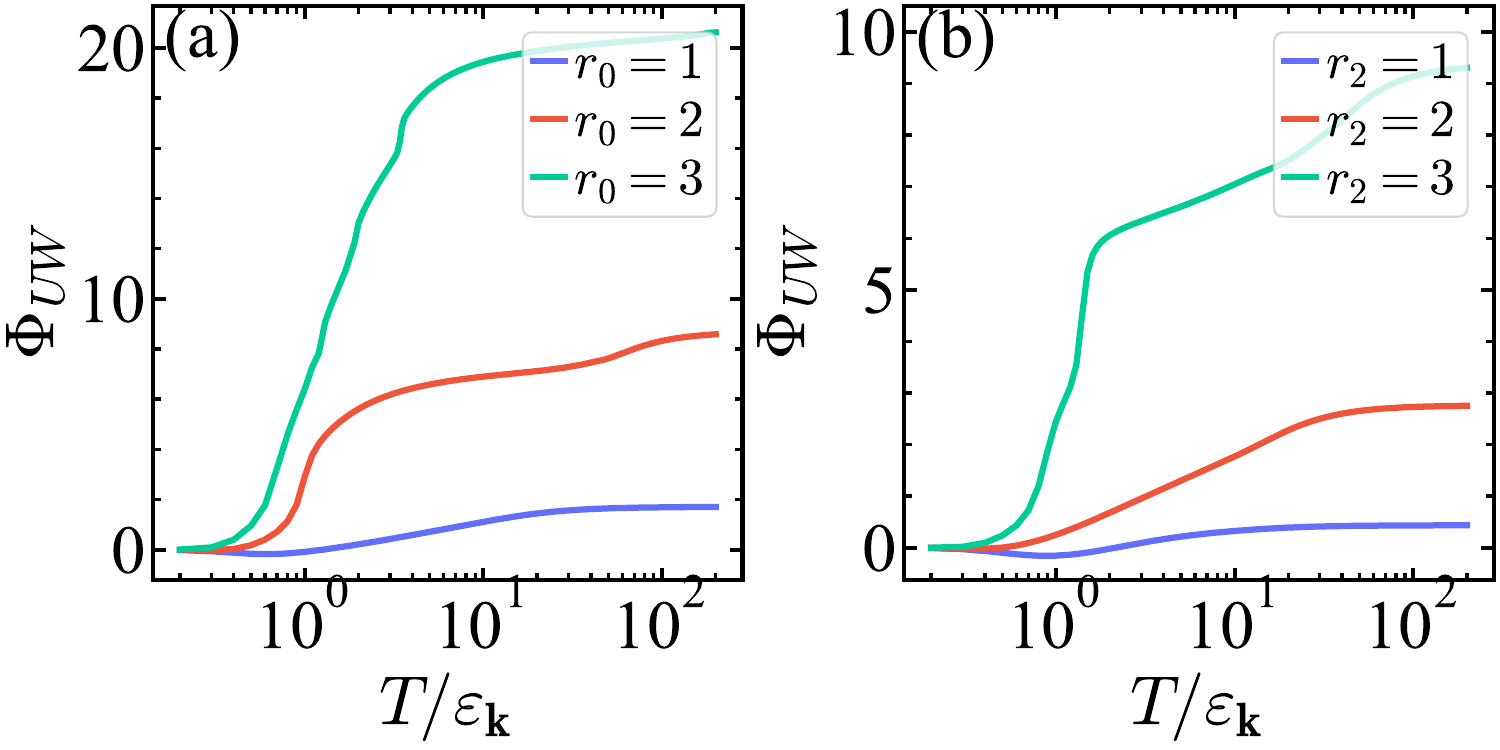}
  \caption{The Uhlmann phase winding $\Phi_{UW}(T)$ varying as the temperature $T$ increases for (a) the path shown in Figure \ref{paths}(a) and (b) for the path shown in Figure \ref{paths}(b). Note that the x-axis is in logarithmic scale.}
  \label{fig:phi_uw_t}
\end{figure}

On the other hand, one may wonder whether it is meaningful to consider the Uhlmann phase as $T\to \infty$, since the BEC collapses when the temperature is above the transition temperature $T_c$, and the Bogoliubov approximation becomes invalid. To address this concern, we show the Uhlmann phase winding $\Phi_{UW}(T)$ with varying $T$ in Figure \ref{fig:phi_uw_t} for the two kinds of paths in Figure \ref{paths} respectively. These results suggest that the Uhlmann phase winding $\Phi_{UW}(T)$ mostly increases in low temperature. Our numerical results indicate that when the temperature rises to approximately $20\varepsilon_\bk$, the Uhlmann phase winding $\Phi_{UW}(T)$ reaches about $90\%$ of its total value $\Phi_{UW}(T\to\infty)$. Consequently, we anticipate that the Uhlmann winding occurs when the quasiparticle energy $\varepsilon_\bk$ is much lower than the critical temperature $T_c$. This condition can be met by choosing an appropriate momentum.

The Uhlmann phase is applicable not only to finite-temperature systems but also to mixed states in open quantum systems. And the density matrix at infinite temperature can also describe the maximally mixed states in open quantum system. While it is known that maximally mixed states always have zero Uhlmann phase, the converse is not necessarily true; a density matrix that differs from the identity matrix can also yield a vanishing Uhlmann phase. This observation raises interesting questions for future research, such as whether Uhlmann phase winding occurs in open quantum systems, and if a Uhlmann phase winding number can be meaningfully extracted and interpreted for such systems.

\section{Proposal for experimental measurement of Uhlmann phase of BEC}\label{sec:experiment}

The experimental implementations to measure Uhlmann phase have been proposed for various systems, such as the photonic systems \cite{ericsson_mixed_2003}, and spin systems \cite{hou_finitetemperature_2021}. Recently, in \cite{viyuela_observation_2018}, the Uhlmann phase has been measured in topological system by using an IBM's quantum computing platform. In these cases, the experimental schemes to measure Uhlmann phase are similar, i.e. based on a purification and interference scheme. Firstly, one needs to find a purification $\ket{w(0)}$ of the density matrix $\rho_0$. Here, $\ket{w(0)}$ is a state in the Fock space $\mathbb{F}_S\otimes \mathbb{F}_A$, where $\mathbb{F}_S$ is the Fock space of the physical system $S$ and $\mathbb{F}_A$  denotes the Fock space of the ancilla system $A$. Here, the ancilla system is another BEC. By doing partial trace over the ancilla system, one obtains the original density matrix $\rho_0 = \mathrm{Tr}_A \ket{w(0)}\bra{w(0)}$. Secondly, one needs to find a time evolution process for the purified state $\ket{w(t)} = U(t)\ket{w(0)}$, such that $\mathrm{Arg}\langle w(0)|w(1)\rangle = \mathrm{Arg}(\rho_0 \mathcal{P} e^{\int A_U dt}) = \Phi_U$. Finally, one can use an interferometer device, in the BEC case the atomic interferometer, to measure the Uhlmann phase $\Phi_U$ \cite{cronin_optics_2009}. A similar interference measurement method has been used in the so-called $SU(1,1)$ interferometer \cite{gabbrielli_spinmixing_2015, marino_effect_2012, yurke_su_1986}.

We start from finding the purification of the density matrix $\rho(t) = \sum_\bf{n} \lambda_\bf{n} \ket{\Psi_\bf{n}(t)}\bra{\Psi_\bf{n}(t)}$ at parameter $t$. The ancilla system we choose here is another BEC. A simple purified state is $\ket{w_0(t)} = \sum_\bf{n} \sqrt{\lambda_\bf{n}} \ket{\Psi_\bf{n}(t)} \otimes \ket{\tilde{\Psi}_\bf{n}(t)}$, where $\ket{\tilde{\Psi}_\bf{n}}$'s are exactly the same quasi-particles state as $\ket{\Psi_\bf{n}}$ but live in the Fock space $\mathbb{F}_A$ of the ancilla system. However, such a purification is not unique, any unitary transformation of the ancilla system 
\begin{eqnarray}
  \ket{w(t)} = \sum_\bf{n} \sqrt{\lambda_\bf{n}} \ket{\Psi_\bf{n}(t)} \otimes \tilde{U}_g(t)\ket{\tilde{\Psi}_\bf{n}(t)}
\end{eqnarray} 
can give another purification of the density matrix. This gives us a gauge redundancy. In order to make $\mathrm{Arg}\langle w(0)|w(1)\rangle$ gives us the correct Uhlmann phase, we need to properly choose the gauge. Such a gauge can be identified by imposing the so-called parallel transport condition, i.e. \cite{viyuela_symmetryprotected_2015}
\begin{eqnarray}
  \rm{Im}\bra{w(t)}\frac{d}{d t}\ket{w(t)} = 0.
\end{eqnarray}
Actually, this condition can be satisfied by choose the gauge as $\tilde{U}_g(t)(t) = \mathcal{P}e^{\int_0^t A_U(t') dt'}$ \cite{viyuela_symmetryprotected_2015}.

On the other hand, the quasi-particle state $\ket{\Psi_\bf{n}(t)}$ can also be obtained by time evolution of $\ket{\Psi_\bf{n}(0)}$ under a suitable chosen Hamiltonian, i.e. $\ket{\Psi_\bf{n}(t)} = U_d(t) \ket{\Psi_\bf{n}(0)}$. Hence, we have
\begin{eqnarray}
  \ket{w(t)} = \sum_\bn \sqrt{\lambda_\bn} U_d(t) \ket{\Psi_\bf{n}(0)} \otimes [\tilde{U}_g(t) \tilde{U}_d(t)] \ket{\tilde{\Psi}_\bf{n}(0)}.
\end{eqnarray}
Then by using Eq.(\ref{dn}), we can rewrite the above equation as 
\begin{eqnarray}
  \ket{w(t)} &=& \sum_\bn \sqrt{\lambda_\bn} U_d(t)D(\zeta_0)\ket{\bn}_S\otimes \tilde{U}_g(t)\tilde{U}_d(t)\tilde{D}(\zeta_0) \ket{\bn}_A \nn\\
   &\equiv& U_d(t)D(\zeta_0)\tilde{U}_g(t)\tilde{U}_d(t)\tilde{D}(\zeta_0) \ket{m},
\end{eqnarray}
where we have defined $\ket{m} = \sum_\bn \sqrt{\lambda_\bn} \ket{\bn}_S\otimes \ket{\bn}_A$. Actually, the state $\ket{m}$ is nothing but the coherent state of a two-component BEC, if we view the bosonic modes in the physical system and the ones in the ancilla system as two species of bosons in the two-component BEC. More specifically, we can define the vacuum of the composite system as $\ket{0} = \ket{0}_S\otimes \ket{0}_A$. Then the state $\ket{m}$ can be written as 
\begin{eqnarray}
  \ket{m} &=& \sum_\bn \sqrt{\lambda_\bn} \frac{\ak^{\dag n_1}\amk^{\dag n_2}}{\sqrt{n_1!n_2!}} \frac{\tilde{a}_{\bb}^{\dag n_1}\tilde{a}_{-\bb}^{\dag n_2}}{\sqrt{n_1!n_2!}}\ket{0} \nn \\
   &=& \frac{1}{\sqrt{\mathcal{Z}}}\sum_\bn \frac{(e^{-\frac{\varepsilon_\bb}{2T}}\ak^\dag \tilde{a}_\bb^\dag)^{n_1}}{n_1!} \frac{(e^{-\frac{\varepsilon_\bb}{2T}}\amk^\dag \tilde{a}_{-\bb}^\dag)^{n_2}}{n_2!} \ket{0} \nn\\
   &=& \frac{1}{\sqrt{\mathcal{Z}}}\exp(e^{-\frac{\varepsilon_\bb}{2T}}\ak^\dag \tilde{a}_\bb^\dag + e^{-\frac{\varepsilon_\bb}{2T}}\amk^\dag \tilde{a}_{-\bb}^\dag) \ket{0},
\end{eqnarray}
which is a coherent state of two-component BEC. Here, we have used $\lambda_\bn = e^{-(n_1 + n_2)\varepsilon_\bb/T}/\mathcal{Z}$. Using similar technique as in Eq.(\ref{d_op}), we can rewrite $\ket{m}$ as time evolution of the vacuum state under Hamiltonian $H_m$
\begin{eqnarray}
  \ket{m} = e^{-i H_{m}}\ket{0},
\end{eqnarray}
where $H_m = i\zeta_m(\ak^\dag \tilde{a}_\bb^\dag + \amk^\dag \tilde{a}_{-\bb}^\dag) + h.c.$, with $\zeta_m = \arctan(e^{-\varepsilon_\bb/2T})$.

Thus, one can prepare the initial state $\ket{w(0)}$ as follows: starting from the fully condensed state of two-component BEC, i.e. all particles condensed in zero momentum states, and quench the system under the Hamiltonian $H_m$ to get $\ket{m}$, then change quenched Hamiltonian to $H_{\zeta_0} = \zeta_0(\ak^\dag \amk^\dag + \tilde{a}_\bb^\dag \tilde{a}_{-\bb}^\dag) + h.c.$, since it can be verified that $e^{-iH_{\zeta_0}} = D(\zeta_0)\tilde{D}(\zeta_0)$, i.e.
\begin{eqnarray}
  \ket{w(0)} = e^{-iH_{\zeta_0}}\ket{m}.
\end{eqnarray}
After this two quench processes, one can obtain the $\ket{w(0)}$ state. 

\begin{figure}
  \includegraphics[width = .45\textwidth]{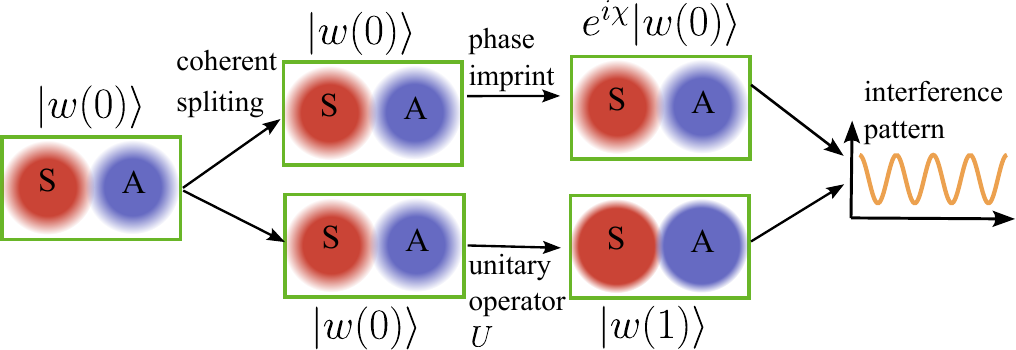}
  \caption{Schematics for the interferometric measurement process. After preparing $\ket{w(0)}$ state, we coherently split the whole system into two copies. One copy is imprinted a phase $e^{i\chi}$ by applying laser beam. The other copy is evolved to $\ket{w(1)}$ state. Then these two copies are brought together to interfere.}
  \label{fig:exp}
\end{figure}

In order to measure the Uhlmann phase in an atomic interferometer, one needs to coherently split the system to two copies (similar to the two arms of Mach-Zehnder interferometer).  In Fig. \ref{fig:exp}, we show the schematics for the interferometric process. Then, making one copy evolve under the time evolution operator $U = U_d(1)\tilde{U}_g(1)\tilde{U}_d(1)$ to obtain $\ket{w(1)}$, i.e.
\begin{eqnarray}
  \ket{w(1)} = U\ket{w(0)}\equiv U_d(1)\tilde{U}_g(1)\tilde{U}_d(1) \ket{w(0)}.
\end{eqnarray}
For the other copy, one needs to imprint a phase factor $e^{i\chi}$ on it by using laser beam \cite{pethick_bose_2008}.The phase imprinting technique has been experimentally implemented in BEC system. \cite{burger_dark_1999, denschlag_generating_2000} Finally, one combine the two arms of the atomic interferometer and measure the interference pattern 
\begin{eqnarray}
&& \Big|\frac{1}{\sqrt{2}}(e^{i\chi}\ket{w(0)} + \ket{w(1)})\Big|^2 \nonumber\\
&&\quad = 1 + |\langle w(0)|w(1)\rangle| \cos(\chi + \Phi_U).
\end{eqnarray}
Thus, by varying $\chi$ and repeating this interferometric measurement process, one can obtain the Uhlmann phase from the interference pattern.

\section{Conclusion}\label{sec:conclusion}

In this paper, we have investigated the Uhlmann phase for Bose-Einstein condensates at finite temperature. By exploiting the $SU(1,1)$ symmetry of the Bogoliubov Hamiltonian, we derived a general formula for the Uhlmann phase of BECs. Our numerical calculations revealed that, contrary to previous studies, the Uhlmann phase can differ from the Berry phase in the zero-temperature limit. Furthermore, we uncovered a unique winding behavior of the Uhlmann phase as the temperature increases, where the total winding degree is equal to the negative of the Berry phase at zero temperature. This winding indicates that the Uhlmann phase takes values on a Riemann surface, highlighting its geometric nature.

Our work not only provides a theoretical framework for understanding topological properties of BECs at finite temperature but also proposes an experimental scheme to measure the Uhlmann phase using an atomic interferometer. These results reveal new aspects of the Uhlmann phase and its relationship with the Berry phase in quantum systems at finite temperature, suggesting that the Uhlmann phase can serve as a useful tool to probe topological properties of BECs and other quantum systems.

Future research directions include extending our analysis to two-component or spinor BECs and exploring the effects of interactions and symmetry breaking on the Uhlmann phase. Additionally, studying the Uhlmann phase of BECs in the presence of noise, decoherence, and measurement errors can provide insights into its robustness and sensitivity to these factors.

We anticipate that our work will stimulate further interest and research on the Uhlmann phase and its applications in quantum physics, particularly in the context of topological properties and experimental investigations of mixed-state quantum systems.

\begin{acknowledgements}
CYW is supported by the Shuimu Tsinghua scholar program at Tsinghua University. YH is supported by the National Natural Science Foundation of China under Grant No. 11874272.

\end{acknowledgements}

\begin{widetext}
\appendix

\section{Detailed derivation for the expression of the Uhlmann connection Eq.(\ref{eq-AU})}\label{app:detail}

In this appendix, we provide a detailed derivation of the Uhlmann connection Eq.(\ref{eq-AU}), following the approach in Ref.\cite{viyuela_symmetryprotected_2015}. 

For an adiabatic path with associated density matrix $\rho(t)$, we seek a decomposition of the form $\rho(t) = w(t)w^\dag(t)$ with $w(t) = \sqrt{\rho}W(t)$, such that adjacent amplitudes $w(t)$ and $w(t+dt)$ satisfy the parallel condition in Eq.(\ref{para}):
\begin{eqnarray}\label{wwdag}
w^\dag(t)w(t+dt) = w^\dag(t+dt)w(t) > 0.
\end{eqnarray}
According to Eq.(\ref{eq-UU}), to satisfy the parallel condition, the amplitude should fulfill:
\begin{eqnarray}
W(t+dt)W^\dag(t) = \sqrt{\rho_2^{-1}}\sqrt{\rho_1^{-1}}\sqrt{\sqrt{\rho_1}\rho_2\sqrt{\rho_1}},
\end{eqnarray}
where $\rho_1 = \rho(t)$ and $\rho_2 = \rho(t+dt)$. By introducing infinitesimal changes $W(t+dt) = W(t) + dW \equiv W(t) + \dot{W}(t)dt$ and $\rho(t+dt) = \rho(t) + d\rho \equiv \rho(t) + \dot{\rho}(t)dt$, where $\dot{W}(t) = dW/dt$ and $\dot{\rho}(t) = d\rho/dt$, Eq.(\ref{wwdag}) can then be rewritten as:
\begin{eqnarray}
(W + dW)W^\dag = (\sqrt{\rho} + d\sqrt{\rho})^{-1}\sqrt{\rho^{-1}}\sqrt{\sqrt{\rho}(\rho + d\rho)\sqrt{\rho}},
\end{eqnarray}
where we've used $\sqrt{\rho + d\rho} = \sqrt{\rho} + d\sqrt{\rho}$. We introduce an auxiliary real parameter $s$ to each differential in the above expression:
\begin{eqnarray}
(W + sdW)W^\dag = (\sqrt{\rho} + s d\sqrt{\rho})^{-1}\sqrt{\rho^{-1}}\sqrt{\sqrt{\rho}(\rho + s d\rho)\sqrt{\rho}}.
\end{eqnarray}
Expanding this expression around $s = 0$ and keeping only first-order terms in $s$, we obtain the Uhlmann connection:
\begin{eqnarray}\label{adt}
Adt = dWW^\dag = \left[\frac{d}{ds}(\sqrt{\rho}+sd\sqrt{\rho})^{-1}\right]\bigg|_{s=0}\sqrt{\rho} + \rho^{-1}\left[\frac{d}{ds}\sqrt{\sqrt{\rho}(\rho + s d\rho)\sqrt{\rho}}\right]\bigg|_{s=0}.
\end{eqnarray}
For the first term on the right-hand side, to the leading order in $s$:
\begin{eqnarray}
(\sqrt{\rho} + sd\sqrt{\rho})^{-1} = \left[\sqrt{\rho}(1 + s\sqrt{\rho^{-1}}d\sqrt{\rho})\right]^{-1} = (1 - s\sqrt{\rho^{-1}}d\sqrt{\rho})\sqrt{\rho^{-1}}.
\end{eqnarray}
Thus,
\begin{eqnarray}
\left[\frac{d}{ds}(\sqrt{\rho}+sd\sqrt{\rho})^{-1}\right]\bigg|_{s=0}\sqrt{\rho} = -\sqrt{\rho^{-1}}d\sqrt{\rho}.
\end{eqnarray}
In the eigenbasis of $\rho$, it becomes
\begin{eqnarray}
-\bra{u_m}\sqrt{\rho^{-1}}d\sqrt{\rho}\ket{u_n} = -\frac{1}{\sqrt{\lambda_m}}\bra{u_m}d\sqrt{\rho}\ket{u_n}.
\end{eqnarray}
We define $K(s)=\sqrt{\sqrt{\rho}(\rho + s d\rho)\sqrt{\rho}}$. Then, the second term in Eq.(\ref{adt}) becomes $\rho^{-1} K'(0)$, where $K'(s) = dK(s)/ds$. Taking the derivative of $[K(s)]^2$ with respect to $s$:
\begin{eqnarray}
K'(s)K(s) + K(s)K'(s) = \sqrt{\rho}d\rho \sqrt{\rho}.
\end{eqnarray}
At $s = 0$, noting that $K(0) = \rho$:
\begin{eqnarray}
K'(0)\rho + \rho K'(0) = \sqrt{\rho}d\rho \sqrt{\rho}.
\end{eqnarray}
Taking matrix elements in the eigenbasis of $\rho$:
\begin{eqnarray}
(\lambda_m + \lambda_n)\bra{u_m}K'(0)\ket{u_n} = \sqrt{\lambda_m \lambda_n}\bra{u_m}d\rho\ket{u_n}.
\end{eqnarray}
Thus, $\rho^{-1}K'(0)$ becomes:
\begin{eqnarray}
\bra{u_m}\rho^{-1}K'(0)\ket{u_n} = \frac{\sqrt{\lambda_n}}{\sqrt{\lambda_m}(\lambda_m + \lambda_n)}\bra{u_m}d\rho\ket{u_n}.
\end{eqnarray}
Noting that $d\rho = d(\sqrt{\rho}\sqrt{\rho}) = d\sqrt{\rho}\sqrt{\rho} + \sqrt{\rho}d\sqrt{\rho}$:
\begin{eqnarray}
\bra{u_m}\rho^{-1}K'(0)\ket{u_n} = \frac{\sqrt{\lambda_n}(\sqrt{\lambda_m} + \sqrt{\lambda_n})}{\sqrt{\lambda_m}(\lambda_m + \lambda_n)} \bra{u_m}d\sqrt{\rho}\ket{u_n}.
\end{eqnarray}
Combining these results, the Uhlmann connection becomes:
\begin{eqnarray}
\bra{u_m}Adt\ket{u_n} = \left[-\frac{1}{\sqrt{\lambda_m}} + \frac{\sqrt{\lambda_n}(\sqrt{\lambda_m} + \sqrt{\lambda_n})}{\sqrt{\lambda_m}(\lambda_m + \lambda_n)}\right] \bra{u_m}d\sqrt{\rho}\ket{u_n} = \frac{\sqrt{\lambda_n} - \sqrt{\lambda_m}}{\lambda_m + \lambda_n} \bra{u_m}d\sqrt{\rho}\ket{u_n}.
\end{eqnarray}
Substituting $\sqrt{\rho} = \sum_i \sqrt{\lambda_i} \ket{u_i}\bra{u_i}$ to above expression
\begin{eqnarray}
\bra{u_m}Adt\ket{u_n} = \frac{\sqrt{\lambda_n} - \sqrt{\lambda_m}}{\lambda_m + \lambda_n}\left[d\sqrt{\lambda_m} \delta_{mn} + \sqrt{\lambda_n}\bra{u_m}d\ket{u_n} + \sqrt{\lambda_m}\left(d\bra{u_m}\right)\ket{u_n}\right],
\end{eqnarray}
where $\delta_{mn}$ is the Kronecker delta. Since $d(\langle u_m|u_n\rangle) = 0$, we have $\left(d\bra{u_m}\right)\ket{u_n} = -\bra{u_m}d\ket{u_n}$. 

Therefore, the final expression for the Uhlmann connection is:
\begin{eqnarray}
\bra{u_m} A \ket{u_n} = \frac{(\sqrt{\lambda_n} - \sqrt{\lambda_m})^2}{\lambda_m + \lambda_n} \bra{u_m}\frac{d}{dt}\ket{u_n}.
\end{eqnarray}

\section{Calculation of the Uhlmann connection for BEC} \label{app_uhl}
In this appendix, we will present the detailed derivation of the Uhlmann connection of $SU(1,1)$ coherent states Eq.(\ref{a_u}). In order to obtain the Uhlmann connection Eq.(\ref{uhl_con_0}), we first need to calculate $\langle\Psi_\mathbf{n}|\frac{d}{d t}|\Psi_\mathbf{m}\rangle$. By using the expression Eq.(\ref{dn}) for $|\Psi_\mathrm{n}\rangle$ and noticing the Euler decomposition Eq.(\ref{euler}), we have
\begin{eqnarray}
\langle\Psi_\mathbf{n}|\frac{d}{d t}|\Psi_\mathbf{m}\rangle &=& \langle \mathbf{n}| D^\dag(\zeta) \frac{d}{d t}D(\zeta) |\mathbf{m}\rangle \nn\\
&=& ie^{i\frac{n-m}{2}\theta} \langle\mathbf{n}|\Big(\dot{\theta} e^{irK_2}K_0 e^{-irK_2} - \dot{r}K_2 -n\dot{\theta}\Big)|\mathbf{m}\rangle \nn\\
&=& ie^{i\frac{n-m}{2}\theta} \langle\mathbf{n}|\Big(\dot{\theta} \cosh r K_0 - \sinh r K_1 - \dot{r}K_2 -n\dot{\theta}\Big)|\mathbf{m}\rangle,
\end{eqnarray}
Here $n=n_1+n_2$ and $m=m_1+m_2$. In the last step we have used
\be
e^{irK_2}K_0 e^{-irK_2} = \cosh r K_0 - \sinh r K_1.
\ee
Then substituting above expression to Eq.(\ref{uhl_con_0}), and noticing the summation is over $\mathbf{m}\neq \mathbf{n}$, we find that
\begin{eqnarray}
  A_U =i \sum_{\mathbf{n}\neq\mathbf{m}}f_{\mathbf{n},\mathbf{m}}
  e^{i\theta K_0}e^{-irK_2}|\mathbf{n}\rangle \langle \mathbf{n}|\Big(-\dot{\theta}\sinh r K_1 - \dot{r} K_2\Big)|\mathbf{m}\rangle\langle \mathbf{m}|e^{irK_2}e^{-i\theta K_0}.
\end{eqnarray}
Here we defined $f_{\mathbf{n},\mathbf{m}}=\frac{(\sqrt{\lambda_\mathbf{n}} - \sqrt{\lambda_\mathbf{m}})^2}{\lambda_\mathbf{n} + \lambda_\mathbf{m}}$. 
Noticing that $K_1 = (K^+ + K^-)/2,\ K_2 = (K^+ - K^-)/2i$, thus, only terms with $\mathbf{m} = \mathbf{n}\pm (1,1)$ contribute non-zero values, other terms vanish. 

Also, when $\mathbf{m} = \mathbf{n}\pm (1,1)$, we have $f_{\mathbf{n},\mathbf{m}}=1-1/\cosh(\beta\varepsilon_\bk)\equiv\eta$ which is a constant and can be pulled out the summations. Therefore, we can make use of the completeness relation to simplify $A_U$ as follows
\begin{eqnarray}
  A_U &=&-i\eta \sum_\mathbf{n}  e^{i\theta K_0}e^{-irK_2} (\dot{\theta}\sinh r K_1 + \dot{r} K_2) 
  |\mathbf{n}\rangle\langle \mathbf{n}|e^{irK_2} e^{-i\theta K_0} \nn\\
   &=& -i\eta \,e^{i\theta K_0}e^{-irK_2} (\dot{\theta}\sinh r K_1 + \dot{r} K_2) e^{irK_2} e^{-i\theta K_0},
\end{eqnarray}
Then by using the transformation relations \cite{puri_mathematical_2001}
\begin{eqnarray}
  e^{i\theta K_0}K_2 e^{-i\theta K_0} &=& \sin\theta K_1 + \cos\theta K_2, \\
  e^{-irK_2}K_1 e^{irK_2} &=& \sinh r K_0 + \cosh r K_1, \\
  e^{i\theta K_0}K_1 e^{-i\theta K_0} &=&  \cos\theta K_1 - \sin\theta K_2,
\end{eqnarray}
we can obtain Eq.(\ref{a_u}).

\end{widetext}

\bibliography{ref}

\end{document}